\documentclass[
    ,final            
  ,numberedheadings 
  ]
  {aipproc}

\layoutstyle{6x9}

\begin{document}

\title{Diagnostics
of Plasma Properties in  Broad  Line Region of AGNs}

\author{L. \v C. Popovi\'c}{
address={Astronomical Observatory,  Volgina 7, 11160
Belgrade
74, Serbia (lpopovic@aob.bg.ac.yu)}
  ,address={Astrophysikalisches Institut Potsdam, An der Sternwarte 1,
D-14482
Potsdam, Germany (Alexander von Humboldt Fellow)}
}

\begin{abstract}
The Boltzmann-plot (BPT) method for laboratory plasma diagnostic was used
for a
quick estimate of physical conditions in the
 Broad Line Region (BLR) of 14 Active Galactic Nuclei   
(AGNs). For the  BLR of nine AGNs, where PLTE exist, the estimated
electron temperature are in the range T$\sim$ 130000K - 37000K are in good
agreement with previous estimates. The estimated electron densities depend
on the opacity of the emitting plasma in the BLR, and they are in range
from $10^9\rm cm^{-3}$ (for optically thick plasma) to $10^{14}\rm
cm^{-3}$  (for optically thin plasma). Although an alternative of PLTE in
some AGNs may be very high intrinsic reddening effect, this method may be
used for fast insight into physical processes in the BLR prior to applying
more sophisticated physical models.
\end{abstract}

\maketitle


\section{Introduction}

The photoionization, recombination and collisions can be considered
as relevant
processes in Broad Line Region (BLR) of Active Galactic Nuclei (AGNs). At
larger ionization parameters recombination is
more important, but at the higher temperatures the collisional excitation
also becomes important, as well as in the case of low ionization
 parameters.
These two effects, as well as radiative-transfer effects in Balmer lines  
should be taken into account to explain the
 ratios of Hydrogen lines  (Osterbrock 1989, Krolik 1999).
Different  physical conditions and processes can be
assumed in order to use the emission lines for diagnostic of emission
plasma
(Osterbrock 1989, Griem 1997, Ferland et al. 1998). Although "in
many
aspects the BLRs are physically as  closely related to stellar atmospheres
as traditional nebula" (Osterbrock  1989), the plasma in the  BLR
probably
does  not come close to being in complete Local Thermodynamical
Equilibrium (LTE).
 However, there may still be the Partial Local Thermodynamical
Equilibrium (PLTE)
in the sense that  populations of sufficiently highly excited levels are
related to the next  ion's ground state population by Saha-Boltzmann
relations (van der Mullen  et al. 1994), or to the total
population in all fine-structure levels of the ground-state configuration 
(see e.g. Griem 1997). The PLTE for different types of plasma: ionizing,
recombining plasma and plasma in ionization balance were discussed in
Fujimoto \& McWhirter (1990). They found that the populations of
higher-lying levels are well described by the Saha-Boltzmann equation.
Recently, Popovi\'c et al. (2002) found  that the Balmer lines of
NGC 3516 indicate that the Balmer emitting line  region  may be in the
PLTE.

Here we present a part of the results that will be given in more details  
in Popovi\'c
(2003) concerning the test of the PLTE existence in BLR of AGNs  using 
the Boltzmann-plot (BPT) of Balmer lines. We 
discuss the possibility of estimation of the relevant  physical processes
and plasma parameters  in BLRs using this method.

\section{Theoretical remarks}

If the plasma is in PLTE, the population
of the parent energy  states adhere to a Boltzmann distribution
uniquely
characterized by
their excitation temperature ($T_e$), and this temperature may
be
obtained from a Boltzmann-plot when the transitions within the same  
spectral series are considered (for more details see Konjevi\'c 1999,
Popovi\'c 2003)

 $$\log(I_n)=\log{I_{ul}\cdot
\lambda\over{g_uA_{ul}}}=B-A{E_u},\eqno(1)$$
where $I_{lu}$ is relative intensity of transition from upper to lower
level ($u\to l$), $B$ and $A$
are constants, where $A$ indicates temperature and we will
call it the temperature parameter.
  
If we can approximate the $\log(I_n)$ as a linear decreasing function
of
$E_u$
then: a) it indicates that PLTE may exist  at least to some extent
in the BLR;
b) if  PLTE is present, the
population adhering to a Boltzmann distribution is uniquely characterized
by its excitation temperature. Then we can estimate the electron
temperature from Eq. (1), $T_e=1/(kA)$, where $k=8.6171\cdot
10^{-5}\rm eV/K$ is the Boltzmann constant;
c) if PLTE is present  we can roughly estimate the minimal electron  
density in BLR.
 Here, we should mention that "Case B" recombination
of Balmer lines  can bring the $\log(I_n)$ {\it vs} $E_u$
as linear decreasing function (Osterbrock 1989). But, regarding the
physical
conditions
(electron densities and temperatures) in BLRs in this case
 the constant A is too small ($A<0.2$) and the Boltzmann-plot cannot
be
applied
for diagnostics of electron temperature even if PLTE exists  (see
discussion in Popovi\'c 2003).
Moreover, in this case Boltzmann-plot method can be used as
an indicator of "Case B" recombination  in BLR of some AGNs.

\section{Observations and data reduction}

In order to test the existence of PLTE in BLR,
we use HST observations obtained with the Space Telescope Imaging  
Spectrograph (STIS) and Faint Object Spectrograph (FOS),  covering the
wavelength ranges  2900-5700 \AA\ and 6295-6867 \AA\  (rest wavelength).
 From the
very large data base of AGN spectra at HST archive we selected the objects
using following selection criteria:
 a)  the observation covered the Balmer series line wavelength region;  b)
the observations were performed on the same day;
c)  all the lines from Balmer series can be recognized and all have
relatively well defined shapes;
d) we considered only low red-shifted objects.

To perform a test we  subtracted the narrow and  satellite lines from
Balmer lines. To estimate the contribution of these lines we used
a multi-Gaussian analysis (e.g. see Popovi\'c et al. 2001, 2002).
 We fit each line with a sum of Gaussian components using a $\chi^2$
minimalization routine.

On the other hand, the reddening effect can influence  the Balmer lines
ratio (see e.g.  Crenshaw et al. 2001, 2002, and
references therein) and consequently on temperature parameter obtained by
Boltzmann-plot. Here  the  Galactic reddening was taken into account
using the data from NASA's Extragalacitc Database (NED).
  In order to
test the total (Galactic + intrinsic) reddening influence we have
considered the case of Akn 564,
 where the reddening data are given by Crenshaw et al. (2002). We
estimated that the reddening effect can contribute to the Boltzmann-plot
slope
around of 30\%-40\% (in the case of Akn 564 around of 35\%,
$(kA)/(kA)_{redd.}=1.35$), but this effect cannot
qualitatively
disturb the straight line as a function in Boltzmann-plot.
The  reddening effect will
always cause that  temperatures measured by this method will give
smaller values if  this effect is not  taken into
account.
In the rest of our sample we
did not consider the intrinsic reddening
effect.

\begin{figure} 
  \includegraphics[height=.25\textheight]{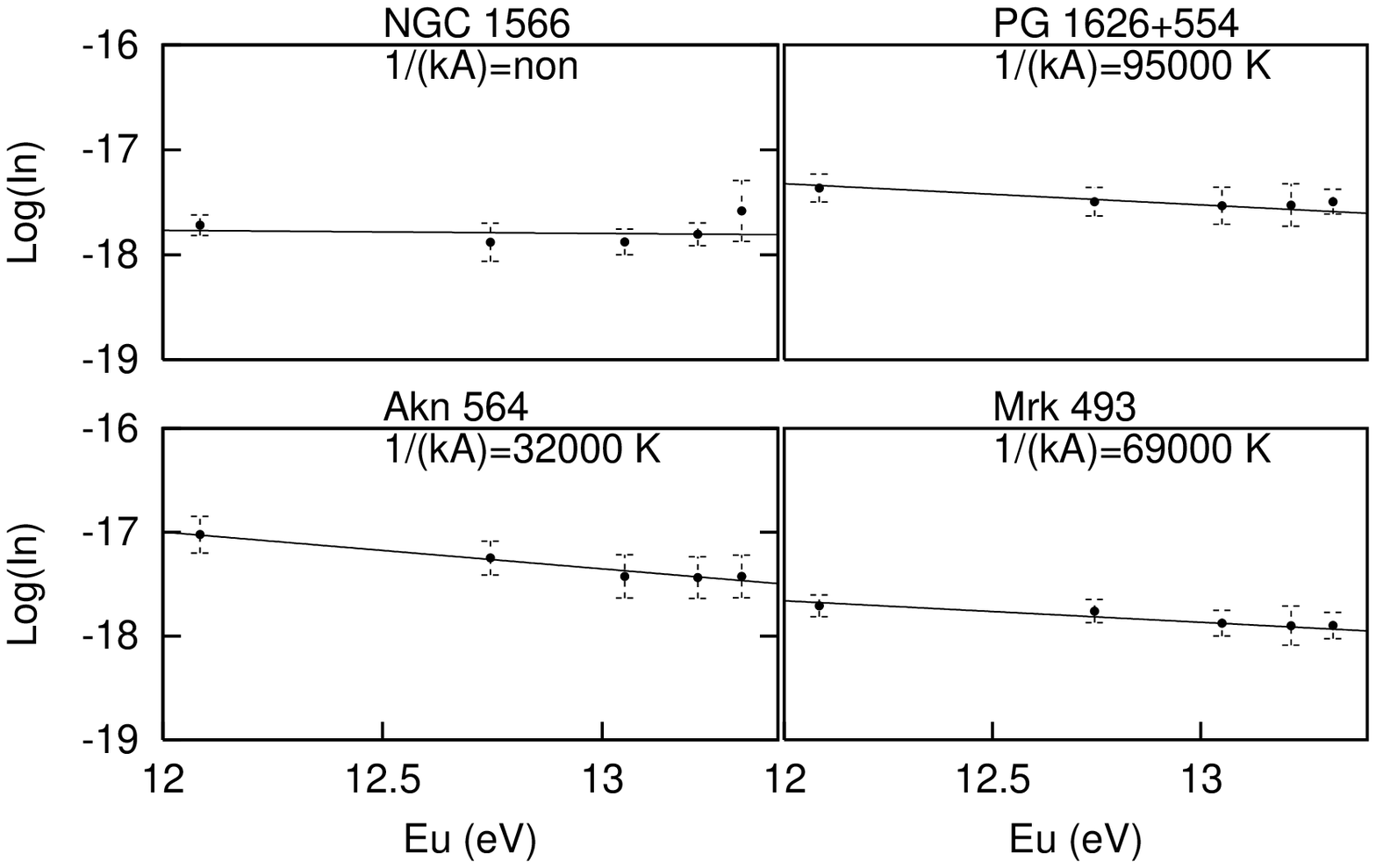}
  \includegraphics[height=.25\textheight]{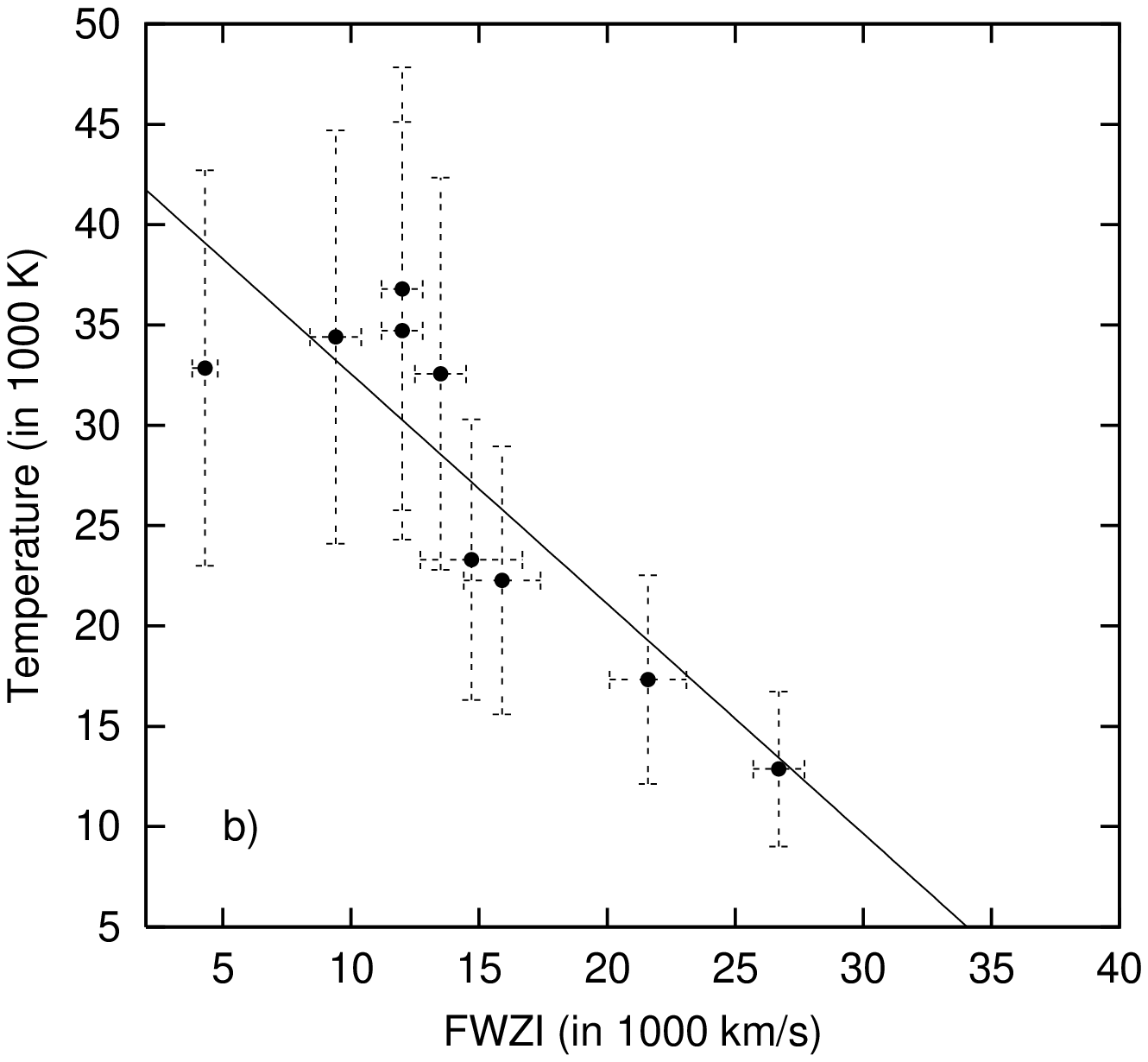}
  \caption{{\it Left}: The Boltzmann-plot for the  Balmer lines of four
AGNs from selected sample (NGC 2566 - BPT does not exist; PG 1626+554 and 
Mrk 493 BPT exists and indicates Case B recombination; Akn 564 - BPT
indicates PLTE). The
corresponding temperature,
or $1/(kA)$, is given at the top of each panel. {\it Right}: Measured
electron temperature as a function FWZI for the AGNs where BPT indicates
presence of PLTE.} \end{figure}

\section{Results and discussion}

The more detailed discussion of the obtained results is given in Popovi\'c
(2003), here we will give the basic conclusion.

From the test we found:

1)  that from the 14 selected AGNs,  in 9 AGNs
Boltzmann-plot indicates the existence of PLTE in BLR,
while in the case of 4  of them   the Boltzmann-plot indicates
"Case B" recombination in BLR. In remaining 1 AGN the Boltzmann-plot
cannot
be applied (see Fig. 1, left).

2) The estimated BLR electron temperatures  using Boltzmann-plot
where PLTE exists are in a range (1.3 - 3.7)$\cdot 10^{4}$ K (within
30\% accuracy). They are in a good agreement with the previous
estimations.

The electron densities in BLR have been considered for optically thin and
optically thick plasma and we found that

i) For optically thin plasma, the electron density in the case of
PLTE, at least in some parts of BLR, should be higher than conventionally
accepted for BLR (order of $10^{14}\rm cm^{-3}$).

ii) For optically thick plasma, the electron density in the case of
PLTE is in  
agreement with the conventionally accepted for BLR ($\sim 10^9\rm
cm^{-3}$).

On the other hand, as one can see in Fig. 1 right, the electron
temperatures
estimated by using
Boltzmann-plot tend to be velocity dependent as a linear decreasing
function of random velocities measured at Full
Width
at Half Maximum (FWHM)  as well as Full Width at Zero Intensity (FWZI).

The Boltzmann-plot method is very useful for probing of the physical
properties of BLRs where PLTE exist. On the other side,
although the  alternative of PLTE indicated by the Boltzmann-plot method
may be considered (e.g. Case B recombination + high intrinsic reddening
effect), the method may be used
for fast insight into  physical processes in BLR of  AGNs prior
applying more sophisticated BLR physical models.


\begin{theacknowledgments}

This work was supported by the Ministry of
Science,
Technologies and Development of Serbia through the project
``Astrophysical Spectroscopy of
Extragalacitc Objects''. Also, the work was supported by Alexander von
Humboldt Foundation through the program for foreign scholars. I  
thank Prof. S. Djeni\v ze, Prof. M.S.
Dimitrijevi\'c and Prof. N. Konjevi\'c for useful discussion and
suggestions.
\end{theacknowledgments}


{}
\end{document}